\documentclass[a4paper]{article}

\usepackage{INTERSPEECH2020}

\usepackage{tipa}
\usepackage{color,pstricks,graphicx}
\usepackage{epstopdf}
\usepackage{multirow}
\usepackage{float}
\usepackage{placeins}

\usepackage{setspace}

\newcommand{\snrninty}{SNR$_{90}$}

\title{Automatic Estimation of Intelligibility Measure for Consonants in Speech}
\name{Ali Abavisani$^1$, Mark Hasegawa-Johnson$^1$}

\address{
  $^{1}$Department of Electrical and Computer Engineering, University of Illinois at Urbana-Champaign}
\email{aliabavi@illinois.edu, jhasegaw@illinois.edu}

\begin{document}

\maketitle

\begin{abstract}

In this article, we provide a model to estimate a real-valued measure of the intelligibility of individual speech segments.
We trained regression 
models based on Convolutional Neural Networks (CNN) for stop consonants \textipa{/p,t,k,b,d,g/} associated with vowel \textipa{/A/}, to estimate the 
corresponding Signal to Noise Ratio (SNR) at which the Consonant-Vowel (CV) sound becomes intelligible for Normal Hearing (NH) ears. The intelligibility 
measure for each sound is called \snrninty, and is defined to be the SNR level at which human participants are able to recognize the consonant
at least 90\% correctly, on average, as 
determined in prior experiments with NH subjects.  
Performance of the CNN is compared to a baseline prediction based on automatic speech recognition (ASR),
specifically, a constant offset subtracted from the SNR at which the ASR becomes capable of correctly labeling the consonant.
Compared to baseline, our models were able to accurately estimate the \snrninty~intelligibility 
measure with less than 2 [dB$^2$] Mean Squared Error (MSE) on average,
while the baseline ASR-defined measure computes \snrninty~with a variance of 5.2 to 26.6 [dB$^2$],
depending on the consonant.

\end{abstract}

\section{Introduction}

The primary purpose of hearing aids is to improve speech perception. But the speech signal has little role in tuning current hearing aid technologies. 
There has been no consensus on how to involve speech in the procedure. In clinical 
audiology, it is usual that hearing impaired (HI) patients complain about their difficulty in understanding speech in noisy environments. A prerequisite 
for proper advice regarding the patient's ability to communicate in noisy situations, or for selection of the optimal hearing aid 
amplification, is a reliable clinical test to assess patient's speech perception in noise. However, developing such a test is a very 
complicated task due to the large number of factors that are involved in the measurements~\cite{Bronkhorst90}. Thus, hearing speech in 
background 
noise should be a substantial part of a clinical audiology test to assess hearing loss (HL). 
If the spectrum of the masker sound is shaped according to the long-term 
average of the speech signal, the test results will be less dependent on the speaker~\cite{Bronkhorst90}.

Psycho-acoustic speech recognition experiments with human subjects using Consonant-Vowel (CV) sounds as speech stimulus have a long 
history~\cite{Miller55},
and can therefore be effectively calibrated. Since about 58\% of the phonetic segments in spoken English are consonants \cite{Minesetal78}, consonant 
recognition scores are appropriate for the evaluation of speech intelligibility.  Recorded CV stimuli vary in their intelligibility, however:
some stimuli that are clearly intelligible under quiet conditions become unintelligible with only a small amount of added noise,
apparently because of the presence of conflicting cues for the place of articulation~\cite{LiMenonAllen10}.
In order to be useful as a test of HL audibility thresholds, it is necessary to 
select CV speech stimuli that are intelligible to NH listeners at the test SNR.

In normal hearing ears each consonant becomes masked at a token dependent
threshold, denoted \snrninty. The \snrninty~is defined as the SNR at which NH ears can recognize the token correctly, with at least 90\% probability,
averaged across NH listeners.
As the noise is increased from Quiet (no noise), the identification of most sounds goes from less than 0.5\% error to 10\% error (at \snrninty), 
and then to chance performance, over an SNR range of just a few [dB] (i.e., less than 10  [dB]) \cite{ToscanoAllen14}. Hence
\snrninty~is an important token-specific threshold metric of noise
robustness. Since \snrninty~is based on NH perception, it is a perceptual measure of understanding speech in noise, for a NH ear. 
It would be very useful if one could assign an \snrninty~label to each speech token so in a speech-based test with background noise, 
the audiologist would know which tokens are appropriate for speech perception assessment. Additionally, as the \snrninty\ is a measure corresponding to 
the intensity of the primary cue region \cite{KapoorAllen12}, knowledge of this perceptual measure could be used to enhance speech playback
in noisy environments: after measuring the background noise level, the playback device could amplify each syllable as necessary to guarantee
that every segment is played at an SNR higher than its own \snrninty.

Previous studies show that, at SNRs well above \snrninty, HI listeners will have errors in recognizing a small subset of CV stimuli out of all the
presented stimuli~\cite{AbavisaniAllen17,TrevinoAllen13b}. Once high error sounds
have been identified, one may seek the optimum treatment (insertion gain) for a patient's hearing aid.

In this study, we propose a model to estimate the \snrninty~for CV speech sounds based on CNN (we use only 1D convolution, 
with an architecture based most closely on \cite{Waibel89a}). The 
model is a supervised estimator of \snrninty~in dB. Particularly, current work is focused on \snrninty~estimation for stop consonants 
\textipa{/p, t, k, b, d, g/} in association with vowel \textipa{/A/}. 
To accomplish this goal, one needs a suitable dataset of CV speech sounds to train the model. One obstacle is that examining each CV 
with 30 NH listeners in psycho-acoustic speech recognition experiments takes a tremendous 
amount of time, and only a handful of sounds can be evaluated this way.  A CNN trained using such a small corpus does not
achieve low error rates.
To overcome this challenge, we propose a speech augmentation method by 
manipulating speech characteristics in ways that do not affect the consonant recognition score for an average NH listener. These manipulations include 
inducing microphone characteristics (high pass or low pass effect with small attenuation), pitch shift (up and down) and consonant duration manipulation 
(compression and extension).

This study is a novel approach to incorporate speech in the process of tuning hearing aids, using machine learning. Previous works 
that use machine learning to assist the design of hearing aids are mainly focused on using deep learning to estimate amplification 
gain \cite{MondolLee19}, or suppress noise and reverberation for speech enhancement \cite{ZhangEtal16}. The models 
proposed here may be used in the audiology clinic to propose perceptual stimuli for hearing aid fine tuning.

The task proposed in this paper is to estimate the SNR at which any particular spoken syllable becomes intelligible to an NH listener.
To the best of our knowledge, there is no published algorithm that performs the same task, therefore there is no published baseline
to which the results of this study can be compared.  In order to create a baseline for comparison, therefore, we use the SNR at which
a commercial ASR (Google cloud's speech to text model \cite{google}) becomes capable of correctly transcribing the same consonant.
Since the threshold SNR for ASR success is always much higher than the threshold SNR for human listeners, our baseline measure is computed
by subtracting a constant offset from the ASR SNR.

Section \ref{expsnr90} explains the psycho-acoustic experiments used to determine the \snrninty~of each token. In section \ref{augmentation}, we describe 
the procedures for speech augmentation and how to generate appropriate labels for distorted sounds. 
The CNN-based model to estimate the intelligibility measure is described in section \ref{model}. 
Results and discussion will follow in section \ref{result}.

\section{\snrninty\ determination}
\label{expsnr90}

To determine the \snrninty~of target tokens, we presented them to NH listeners at various SNRs ranging from -22 to 22 [dB].
Listeners were given 14 buttons, labeled with 14 consonants of English, and were asked to select the consonant they heard.
The speech signal was mixed with speech-weighted noise as described by \cite{PhatakAllen07a} to set the SNR to 
-22. -18, -12, -6, 0, 6, 12, 18 and 22 [dB] respectively. Presentation order was randomized over consonant, talker, and SNR.

The experiment was designed using a two-down-one-up strategy: if the subject recognizes the token correctly, the 
SNR drops two levels [12dB], otherwise it increases one level [6dB]. This schedule is consistent with conventional paradigms in audiology testing. 
If the subject 
loops between two consecutive SNRs at least three times, the presentation concludes for that token.

After collecting the data from all NH listeners, we average the response accuracies for each token at each SNR. For the n$^{\textrm{th}}$ subject 
at each SNR, the probability of correct response for the $c^{\textrm{th}}$ token is calculated as:
\begin{equation}
\label{eq:Pc}
P_c(n,SNR) = \frac{N_{\textrm{correct}}(n,SNR)}{N_{\textrm{total}}(n,SNR)}
\end{equation}
where $N_{\textrm{correct}}(n,SNR)$ is the number of correct recognitions at the specified SNR, and $N_{\textrm{total}}(n,SNR)$ is the total 
number of tokens 
presented at that SNR, for the $n^{\textrm{th}}$ subject. Hence, the average score is:
\begin{equation}
\overline{P_c}(SNR) = \frac{1}{N_{\textrm{sub}}}\sum_{n=1}^{N_{\textrm{sub}}}P_c(n,SNR)
\label{eq:pcsnr}
\end{equation}
where $N_{\textrm{sub}}$ is the number of subjects.

\begin{figure}[htbp!]
\centering
\includegraphics[width=.25\textwidth]{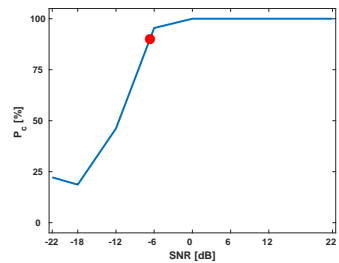}
\caption{An example of determining the \snrninty~of a \textipa{/bA/} token; the \snrninty= -6.66 [dB] is shown with 
a red dot and is determined by linear interpolation along the $P_c$ averaged across 30 NH listeners.}
\label{f:exsnr90}
\end{figure}

The plot of average accuracy ($\overline{P_c}(SNR)$) versus SNR was (in our data) always greater than 90\% at SNRs above \snrninty,
and always less than 90\% at SNRs below \snrninty.  In order to estimate the exact value of \snrninty~for each token, we linearly interpolated
between the smallest $\overline{P_c}(SNR)$ above 90\% and the value of $\overline{P_c}(SNR)$ at the next lower SNR,
and then measured the SNR at which the linear interpolation crosses 90\%; Fig.~\ref{f:exsnr90} shows an example of this procedure.
As can be observed in Fig.~\ref{f:exsnr90},
All NH listeners recognized the \textipa{/bA/} sound whose results are schematized in  Fig.~\ref{f:exsnr90}
with no error above 0 [dB] SNR. At SNR=-6 [dB], 
subjects started to have some errors, but still correctly recognized the CV with accuracy better than 90\%. When the SNR further 
drops to -12 [dB], $\overline{P_c}(SNR)$ suddenly drops below 50\%.
Linear interpolation of $\overline{P_c}(SNR)$ estimates that for this \textipa{/bA/} sound the \snrninty~is -6.66 [dB].

\vspace{-10pt}

Using this procedure, we examined 14 consonants associated with vowel \textipa{/A/}, spoken by 16 different talkers, both male and female 
(total of $14\times16=224$ tokens), 
to determine their \snrninty. To train our model, we selected tokens with \snrninty~below -3 [dB] to focus on tokens with 
better intelligibility measure. To have balanced examples from various talkers, we limited our training 
data to female talkers and stop consonants. Hence, from the pool of 224 evaluated tokens, we selected 39 tokens to build the model 
(8$\times$\textipa{/ka/}, 7$\times$\textipa{/ta/}, 6 of each other stop consonants).
Table \ref{tab:CV} provides the contributing tokens along with their \snrninty~from experiments with NH listeners.

\begin{table}[htbp]
  \caption{\footnotesize CV tokens that are used as original undistorted tokens for training the model to estimate intelligibility measure. 
  For each CV, its \snrninty~in [dB] is provided.}
  \label{tab:CV}
  \centering
  \scriptsize
  \begin{tabular}{l c c c c c c }
    \toprule
    \textbf{Talker} & \textbf{CV} & \textbf{\snrninty} & \textbf{CV} & \textbf{\snrninty} & \textbf{CV} & \textbf{\snrninty} \\
    \midrule
    f101 & /k\textipa{A}/ & -5 & /b\textipa{A}/ & -11 & & \\ \hline
    \multirow{2}{*}{f103} & /p\textipa{A}/ & -17 & /t\textipa{A}/ & -21 & /k\textipa{A}/ & -11 \\
     & /b\textipa{A}/ & -3 & /d\textipa{A}/ & -17 & /g\textipa{A}/ & -13 \\ \hline
    \multirow{2}{*}{f105} & /p\textipa{A}/ & -13 & /t\textipa{A}/ & -17 & /k\textipa{A}/ & -8 \\
	 & /b\textipa{A}/ & -9 & /d\textipa{A}/ & -21 & /g\textipa{A}/ & -11 \\ \hline
    \multirow{2}{*}{f106} & /p\textipa{A}/ & -11 & /t\textipa{A}/ & -11 & /k\textipa{A}/ & -12 \\
     & /g\textipa{A}/ & -5 & & & & \\ \hline
    \multirow{2}{*}{f108} & /p\textipa{A}/ & -11 & /t\textipa{A}/ & -21 & /k\textipa{A}/ & -4 \\
     & /b\textipa{A}/ & -11 & /d\textipa{A}/ & -12 & /g\textipa{A}/ & -3 \\ \hline
    \multirow{2}{*}{f109} & /p\textipa{A}/ & -4 & /t\textipa{A}/ & -17 & /k\textipa{A}/ & -4 \\
     & /b\textipa{A}/ & -11 & /d\textipa{A}/ & -12 & /g\textipa{A}/ & -4 \\ \hline
    \multirow{2}{*}{f113} & /t\textipa{A}/ & -17 & /k\textipa{A}/ & -9 & /b\textipa{A}/ & -4 \\
     & /d\textipa{A}/ & -18 & & & & \\ \hline
    \multirow{2}{*}{f119} & /p\textipa{A}/ & -11 & /t\textipa{A}/ & -18 & /k\textipa{A}/ & -5 \\
     & /d\textipa{A}/ & -18 & /g\textipa{A}/ & -6 & & \\
    
    \bottomrule
  \end{tabular}
  
\end{table}

\vspace{-10pt}
\section{Speech augmentation}
\label{augmentation}


The number of presentations of each CV, during each perceptual experiment, depends on the number of trials necessary to find the SNR at which the
subject makes mistakes (Sec.~\ref{expsnr90}), but on average, each token requires a total listening time of three minutes. 
Each \snrninty~measurement is the result of 30 NH listeners, a total of 90 minutes/token. To train 
a viable CNN-based \snrninty~estimation model for each CV, one would need thousands of different versions of such CV sound, each with a measured \snrninty.
Instead of manually labeling such a large number of training tokens, we began with only 39 labeled tokens, and introduced various 
distortions using methods that have previously been shown to have little effect on the \snrninty~of human listeners.
In this way, it is possible to generate sufficient data to be able to train 
the model.


\subsection{Applied distortions}

The distortions applied in the current study include extending and compressing the duration of the consonant, 
shifting the pitch of the whole CV up and down, and introducing 
channel effects by applying low pass and high pass filtering with small attenuation. We used Praat \cite{praat} for pitch and duration manipulations, 
and MATLAB$^\circledR$ for filtering. Each token (i.e., a CV sound with a specific talker) was distorted according to 
every available single distortion, but not any combinations. Table \ref{tab:dist} provides specifics of distortions applied to each CV sound.

\begin{table}[h]
  \caption{\footnotesize Various distortions applied to each tested CV sound for speech augmentation.}
  \label{tab:dist}
  \centering
  \scriptsize
  \begin{tabular}{l l l}
    \toprule
    \multicolumn{2}{c}{\textbf{Distortion}} & \multicolumn{1}{c}{\textbf{Details}} \\
    \midrule
    \multirow{2}{*}{Duration change} & Extend & from 1:1 up to 1:3 in steps of 0.3\% \\
    						 & Compress & from 1:1 down to 1:0.5 in steps of 1\%  \\ \hline

	\multirow{2}{*}{Pitch Shift} & Up & up to 600 [Hz] in steps of 1 [Hz] \\
    						 & Down &  down to 20 [Hz] in steps of 1 [Hz] \\ \hline
    
	 						 & \multirow{2}{*}{High pass} & \multirow{1}{*}{}20 cut-offs log-spaced between 0.2-3 [kHz], \\
	FIR Filter & & 20 attenuations ranged between 0.6-12 [dB] \\

	(200 degree)   & \multirow{2}{*}{Low pass} & \multirow{1}{*}{}20 cut-offs log-spaced between 1-8 [kHz], \\
    						 & & 20 attenuations ranged between 0.6-12 [dB] \\ 

    \bottomrule
  \end{tabular}
  
\end{table}

The artificially distorted CV token might have a different \snrninty~than the original token. The possible changes in \snrninty~are controlled by 
conducting new psycho-acoustic experiments with NH subjects, similar to the experiment described in section \ref{expsnr90}, to measure the \snrninty~of 
the most distorted token along each distortion continuum. If the measured \snrninty~is greater than 6 [dB], we did not include the 
entire sequence of tokens generated by such distortion in our augmented speech dataset. Table~\ref{tab:notvalid} provides 
the tokens that are not valid after applying various distortions.

\begin{table}[h]
  \caption{\footnotesize CV tokens (vowel \textipa{/A/} is omitted) from various talkers that have \snrninty~greater than 6 [dB] 
  for their most distorted case, thus were \textbf{not} included in the training data.}
  \label{tab:notvalid}
  \centering
  \scriptsize
  \begin{tabular}{l c c c c c c }
    \toprule
    \multirow{2}{*}{\textbf{Talker}} & \multicolumn{2}{c}{\textbf{Duration}} & \multicolumn{2}{c}{\textbf{Pitch Shift}} & \multicolumn{2}{c}{\textbf{Filtering}} \\
    & \textbf{Extend} & \textbf{Compress} & \textbf{Up} & \textbf{Down} & \textbf{LPF} & \textbf{HPF} \\
    \midrule
    f101 & & /k/ & & /k/ & /k/ & /k/ \\
    f103 & /b/ & /k,b,g/ & /b/  &  & /b,g/ &  \\
    f105 & /d/ & /p,k/ & /p,d,g/ &  & /k/ & /p,k,g/ \\
    f106 & /t/ & /t,k/ & & & /t/ & \\
    f108 & /b,d,g/ & /k,g/ & /k,b,g/ & & /k,g/ & /k/ \\
    f109 & /p,g/ & /p,t,k,d,g/ & /k,d,g/ &  & /t,k,g/ &  \\
    f113 & /k,d/ & /k/ & /b/ & & /k/ & \\
    f119 & /p,k,d,g/ & /k,g/ & /k,g/ & /k/ & /k,g/ & /k,g/ \\
    
    \bottomrule
  \end{tabular}
  
\end{table}

For each CV sound, 
we may assume that every token on the continuum, between the original unmodified token and the most distorted 
token, will have an \snrninty~that is somewhere between the original token's \snrninty~and the most distorted token's \snrninty. Thus,
for each distortion scheme,
we linearly interpolated between the \snrninty~of the original and most distorted token
in order to generate \snrninty~labels for tokens in between. Although this procedure 
may not produce the exact experimentally accurate \snrninty~measure for every token, it generates
approximate \snrninty~labels that are useful as training data. 
Using these methods, the original 39 unmodified tokens were expanded to create 43201 tokens (\textipa{/pA/}: 8121 tokens, 
\textipa{/tA/}: 9726 tokens, \textipa{/kA/}: 6267 tokens, \textipa{/bA/}: 7065 tokens, \textipa{/dA/}: 7977 tokens, and \textipa{/gA/}: 4045 tokens).  
Different talkers contributed different token counts for each consonant. This is sufficient CV speech data to train the model for \snrninty~estimation. 


\section{Model structure}
\label{model}

The dataset of unmodified CV tokens contain recorded \textit{wav} files naturally spoken by various talkers (table \ref{tab:CV}). After speech
augmentation, the dataset increased to 43201 \textit{wav} files of CV tokens.
Each file has a duration of less than three seconds 
and contains an isolated CV utterance with a sampling frequency of 16 [kHz]. The data was divided to train, development and test partitions with 
non-overlapping talkers, to train, tune and test the individual model for each stop consonant. 
The percentage of partitions of data was different for various stop consonants, as each 
talker contributed differently in the final dataset after augmentation. The exact number of tokens 
for training, development and test sets are provided in table \ref{tab:NCV}.

\begin{table}[h]
  \caption{\footnotesize Total number of tokens (N$_\textrm{Total}$) after speech augmentation, along with the number of allocated tokens for training 
  (N$_\textrm{Train}$), development (N$_\textrm{Dev}$) and test (N$_\textrm{Test}$), to train models for \snrninty~estimation.}
  \label{tab:NCV}
  \centering
  \scriptsize
  \begin{tabular}{ l c c c c}
    \toprule
    \textbf{CV} & \textbf{N$_{\textrm{Total}}$} & \textbf{N$_{\textrm{Train}}$} & \textbf{N$_{\textrm{Dev}}$} & \textbf{N$_{\textrm{Test}}$} \\
    \midrule
    \textipa{/pA/} & 8121 & 6261 & 1298 & 562  \\
    \textipa{/tA/} & 9726 & 7787 & 974 & 965  \\
    \textipa{/kA/} & 6267 & 4966 & 703 & 598 \\
    \textipa{/bA/} & 7065 & 5622 & 981 & 462  \\
    \textipa{/dA/} & 7977 & 5910 & 1129 & 938   \\
	\textipa{/gA/} & 4045 & 3004 & 569 & 472  \\
    \bottomrule
  \end{tabular}
  
\end{table}

For each CV, the time interval from the start of the consonant till the end of the onset of vowel 
\textipa{/A/} is manually segmented. Within this interval, the 320 point log magnitude Short-Time Fourier 
Transform (STFT) with 75\% overlapping Hamming windows of 
length 25 [msec] is extracted to feed into the input layer.

For each of the sounds \textipa{/pA/}, \textipa{/tA/}, \textipa{/kA/}, \textipa{/bA/}, \textipa{/dA/} and \textipa{/gA/}, we trained a separate 
model to estimate the \snrninty. The models are based on Convolutional Neural Networks (CNN) \cite{Waibel89a}, which include convolutional layers that act 
in the time domain. The model contains 3 to 7 convolutional layers with Rectified Linear Unit (ReLU) nonlinear activation function 
\cite{glorot11ReLU}, reduced to the input of the fully connected layer (FC) by average pooling, 
followed by two fully connected layers with ReLU non-linearity in the hidden layer and a linear output node 
that produces the estimated \snrninty~value. The loss function is Mean 
Squared Error (MSE) between estimated \snrninty~and correct \snrninty~label. We used stochastic gradient descent to minimize the loss. The model is 
implemented in TensorFlow 1.4 \cite{tensorflow}. To avoid over fitting, dropout \cite{srivastava14dropout} 
is applied to the fully connected layer, with dropout rates tuned on the development set.
Table \ref{tab:comnet} illustrates the common structure of the network for various stop consonants. The differences between the models for 
different CVs are in their hyper-parameters.

\vspace{-5pt}
\begin{table}[h]
  \caption{\footnotesize Common structure of the network trained for various CVs.  The number of convolutional layers, and 
    the time domain kernel values of [w$_1$, w$_2$, w$_3$] in convolutional 
    layers, are among the hyper-parameters trained for each CV model separately, and are reported in table \ref{tab:hyp};
    conv4-7 are extra layers added during tuning.}
  \label{tab:comnet}
  \centering
  \scriptsize
  \begin{tabular}{ l c c l l }
    \toprule
    \textbf{Layer} & \textbf{Kernel} & \textbf{(stride, pad)} & \textbf{Input} & \textbf{Output} \\
    \midrule
    conv1 & 1$\times$w$_1$$\times$320$\times$128 & (1,0) & $\log|$STFT$|$ & conv1 \\
	conv2 & 1$\times$w$_2$$\times$128$\times$256 & (1,0) & conv1 & conv2 \\
	conv3 & 1$\times$w$_3$$\times$256$\times$512 & (1,0) & conv2 & conv3 \\
	conv4-7 & 1$\times$w$_3$$\times$512$\times$512 & (1,0) & conv3-6 & conv4-7 \\
	FC & 512$\times$1024  & - & conv3-7 & FC \\
	out & 1024$\times$1  & - & FC & $\hat{\textrm{SNR}}_{\textrm{90}}$ \\
    \bottomrule
  \end{tabular}
  
\end{table}

The hyper-parameters for each model are trained by using the development data. These hyper-parameters include number of convolutional layers,
time domain kernel size for each layer, learning rate for gradient descent optimization, batch size, and dropout rate. 
Table \ref{tab:hyp} provides the parameters for each model after fine-tuning with development data. In table \ref{tab:hyp},
N$_{\textnormal{CNN}}$ indicates the number of convolutional layers, [w$_1$, w$_2$, w$_3$] refers to the time domain kernel size in convolutional layers, 
N$_{\textnormal{Batch}}$ indicates the batch size, and $\eta$ indicates the learning rate. If the network 
has more than three convolutional layers, the time domain kernel size beyond the third layer is set equal to w$_3$.

\begin{table}[h]
  \caption{\footnotesize Hyper-parameters tuned separately for each stop consonant model.}
  \label{tab:hyp}
  \centering
  \scriptsize
  \begin{tabular}{ l c c c l l }
    \toprule
    \textbf{CV} & \textbf{N$_{\textbf{CNN}}$} & \textbf{[w$_\textbf{1}$, w$_\textbf{2}$, w$_\textbf{3}$]} & \textbf{N$_{\textbf{Batch}}$} & \textbf{$\eta$} & \textbf{P$_{\textbf{dropout}}$} \\
    \midrule
    \textipa{/pA/} & 3 & [5, 7, 7] & 8 & $10^{-4}$ & 50\% \\
    \textipa{/tA/} & 7 & [7, 3, 3] & 4 & $10^{-4}$ & 17\% \\
    \textipa{/kA/} & 3 & [7, 5, 7] & 4 & $10^{-5}$ & 38\% \\
    \textipa{/bA/} & 3 & [3, 3, 3] & 16 & $10^{-4}$ & 10\% \\
    \textipa{/dA/} & 7 & [5, 5, 3] & 8 & $10^{-6}$ & 33\% \\
	\textipa{/gA/} & 3 & [5, 3, 7] & 4 & $10^{-6}$ & 8\% \\
    \bottomrule
  \end{tabular}
  
\end{table}

\vspace{-10pt}

\section{Results}
\label{result}

Human speech perception data from experiments with NH listeners form the ground truth for the evaluation of automatic estimates of \snrninty. 
To compare the 
human perception of CV tokens versus machine perception, we tested several commercial ASRs, and chose the one with the best performance for these data,
Google cloud's speech to text interface \cite{google}.
The phone call model in Google's speech to text system is an enhanced model that aims to have better performance in noisy environments.

Since the speech to text system is trained to recognize words and 
sentences, it is unable to recognize non-word CVs, therefore we counted, as correct, 
any word containing the target consonant followed by a non-high vowel.
Output not containing the target CV, and empty output transcript,
were both counted as failure to recognize the CV.

Table \ref{tab:NHvsASR} provides the comparison between human perception of CV sounds versus ASR for stop consonants 
in the test corpus. The lowest SNR at which the 
output transcript of the ASR contained a word including the target CV is reported as the ASR estimate of \snrninty.
The average perceptual evaluation of speech quality (PESQ) score \cite{pesq} is also measured and reported for CV sounds at the ASR \snrninty.
In table \ref{tab:NHvsASR}, the \snrninty s from human and ASR, and the PESQ scores are 
averaged across different talkers (indicated by N$_T$) for the same CV.

\begin{table}[h]
  \caption{\footnotesize \snrninty~(in [dB]) of CV stimuli, average results across N$_T$ talkers, measured by human subjects (ground truth) and ASR. 
  The PESQ score is calulated at ASR \snrninty. Bias (in [dB]) and variance ($\sigma^2$) of ASR \snrninty~estimation 
  (in [dB$^2$]), as well as our model's test MSE (in [dB$^2$]) are provided.}
  \label{tab:NHvsASR}
  \centering
  \scriptsize
  \begin{tabular}{ l l c c c c c c}
    \toprule
    \multirow{2}{*}{\textbf{CV}} & \multirow{2}{*}{\textbf{N$_\textrm{T}$}} & \multicolumn{2}{c}{\textbf{SNR$_{\textbf{90}}$}} & \multirow{2}{*}{\textbf{PESQ}} & \multicolumn{2}{c}{\textbf{ASR}} & \textbf{Our model} \\
     & & \textbf{Human} & \textbf{ASR} &  & \textbf{Bias} & \textbf{$\sigma^\textrm{2}$} & \textbf{Test MSE} \\
    \midrule
    \textipa{/pA/} & 6 & -11.2 & 2.4 & 2.35 & 13.6 & 26.6 & 1.71 \\
    \textipa{/tA/} & 7 & -17.4 & 0.4 & 2.17 & 17.8 & 16.5 & 1.45 \\
    \textipa{/kA/} & 8 & -6.7 & 4.2 & 2.47 & 10.9 & 12.4 & 1.29 \\
    \textipa{/bA/} & 6 & -7.4 & 4.2 & 2.54 & 11.6 & 16.5 & 1.71 \\
    \textipa{/dA/} & 6 & -16.3 & -4.5 & 2.09 & 11.8 & 14.2 & 1.81 \\
	\textipa{/gA/} & 6 & -7 & 4.5 & 2.4 & 11.5 & 5.2 & 1.89 \\
    \bottomrule
  \end{tabular}
  
\end{table}

Variance of the ASR estimated \snrninty~ranged from 5.2 [dB$^2$] (\textipa{/gA/}) to 26.6 [dB$^2$] (\textipa{/pA/}), with an
average of 15.3 [dB$^2$].
In comparison, our CNN-based models were able to 
estimate the \snrninty~of various CV sounds with small errors. The mean squared error of estimation for test sounds were all below 2 [dB$^2$] 
for the models trained for each consonant.
Table \ref{tab:NHvsASR} illustrates the test errors in \snrninty~estimation for stop consonants.

\section{Conclusion}
\label{conclusion}

In this study, we introduced new models based on convolutional neural networks to estimate the \snrninty~of individual speech stimuli.
\snrninty~is defined to be the SNR at  which normal hearing listeners are able to correctly recognize a stimulus with 90\% probability,
and has been shown to be related to the level of the primary cue to consonant identity~\cite{PhatakAllen07a}. 
One important application of such models is to evaluate various speech sounds before using them 
to assess speech perception in humans, e.g., to tune hearing aids.
The main advantage of using the models developed here is to estimate intelligibility of speech syllables in background speech-weighted noise, without 
the need of running expensive and time consuming experiments with human subjects in controlled conditions. The speech augmentation methods introduced 
in the current study help to increase the size of the training database adequately to train deep learning 
models for speech processing. Our results show that 
the developed models outperform the only available baseline, namely, the SNR at which an ASR is able to correctly recognize each consonant.
Variance of the ASR-based estimate of \snrninty~is 15.3 [dB$^2$] (with a bias greather than 11 [dB]), while MSE of the deep-learning-based
estimator is below 2 [dB$^2$].

\section{Acknowledgments}

This work utilizes resources supported by the National Science Foundation’s Major Research Instrumentation program, grant \#1725729, as well as the 
University of Illinois at Urbana-Champaign.

\bibliographystyle{IEEEtran}

\end{document}